\documentclass[12pt,a4paper]{article}

\makeatletter
\@addtoreset{equation}{section}
\makeatother

\begin{document}
\begin{flushright}
\footnotesize
\footnotesize
CERN-TH/2000-025\\
Imperial/TP/99-0/19\\
{\tt hep-th/0001170}\\
January, $2000$
\normalsize
\end{flushright}

\begin{center}

\vspace{.8cm}
{\LARGE {\bf Brane Descent Relations in M-theory}}

\vspace{1cm}

%authors

{\bf Laurent Houart}

\vspace{.1cm}

{
{\it Theory Division, CERN\\
1211 Gen\`eve 23, Switzerland\\
and \\
Theoretical Physics Group\\
Blackett Laboratory,
Imperial College \\
London SW7 2BZ, UK}\\
{\tt l.houart@ic.ac.uk}
}

\vspace{.3cm}

{and}

\vspace{.3cm}

{\bf Yolanda Lozano}

\vspace{.1cm}

{
{\it Theory Division, CERN\\
1211 Gen\`eve 23, Switzerland}\\
{\tt yolanda.lozano@cern.ch}
}

\vspace{.4cm}

\vspace{1cm}

%%%%%%%%%%%%%%%%%%%%%%%%%%%%%%%%%%%%%%%%%%%%%%%%%%%%%%%%%%%%%%%%%%%%%%

{\bf Abstract}

\end{center}
\begin{quotation}

\small

We discuss how the BPS branes of M-theory could be
described as bound states of non-BPS M10-branes.
This conjectured M10-brane is constructed as an unstable spacetime-filling
brane in the massive eleven dimensional supergravity defined with a Killing 
direction, such that 1) the BPS M9-brane is obtained after the tachyonic mode 
of an M2-brane
ending on it condenses, and 2) it
gives the non-BPS D9-brane of the type IIA theory upon reduction. 
The existence of other non-BPS M-branes is also discussed, together
with their possible stabilisation within the Ho\u{r}ava-Witten 
construction.

\end{quotation}

\vspace{1cm}

\newpage

\pagestyle{plain}

\section{Introduction}

Recent progress in the understanding of the role of tachyonic
excitations in unstable brane systems has led to a new
framework in which D-branes appear
as topological defects in the worldvolume of these unstable
systems.  
In this framework the tachyon is considered as a Higgs field which 
tends to develop a stable
vacuum expectation value, and the solitonic configuration that
appears after the condensation is a BPS D-brane (see \cite{reva} 
and references therein). The process can be
iterated by embedding the unstable tachyonic system onto a similar
system of higher dimension, and one can derive in this manner 
brane descent relations.

The mechanism of tachyon condensation can be given a qualitative
description by looking at the couplings in the Wess-Zumino term of
the effective action corresponding to the unstable system. This
analysis allows to describe as well NS-NS charged branes, in particular
fundamental strings, as topological solitons in higher dimensional
systems \cite{Yi,HL1}. In this case
the tachyonic condensing charged object is extended and
non-perturbative, and a more desirable quantitative description is far
out of reach.

In the type IIB theory starting with an unstable system of D9, anti-D9
pairs of branes it is possible to classify the stable Dp-branes 
in the theory by analysing the homotopy groups of the vacuum
manifold of the tachyonic field \cite{Witten}.
Mathematically the Dp-brane charges are
classified by the K-theory groups ${\tilde K}(S^{9-p})$,
which describe the equivalence classes
of pairs of vector bundles that characterise the system of coincident
D9, anti-D9 branes up to creation and annihilation \cite{Witten}. 
S-duality determines that, similarly, NS-NS branes can be 
classified by analysing the
homotopy groups associated to the tachyonic field of systems of
NS9, anti-NS9 pairs of branes \cite{HL2}. 
The K-theory description is identical
to that of D-branes, but the equivalence classes of pairs of vector
bundles characterise instead coincident NS9, anti-NS9 branes.

In the type IIA theory D-brane descent relations can be derived by 
analysing the homotopy groups of the vacuum manifold of the tachyonic 
field in a system of $n$ D8, anti-D8 pairs of branes and, 
recalling that the D8-brane
can be obtained from a non-BPS D9-brane after tachyonic condensation
\cite{Horava}, the Type IIA Dp-branes can finally be described as 
bound states of spacetime-filling branes, in this way preserving all the
symmetries of the ten dimensional spacetime. Mathematically the
D-brane charges are classified by $K^{-1}(S^{9-p})$ K-theory groups
(see \cite{Horava}).

In this paper we analyse the possibility of having 
brane descent relations in M-theory. We show that the brane descent
relations that one expects
to find by oxidation from Type IIA can be
predicted through the analysis of the worldvolume effective action of a
system of $n$ M9, anti-M9 pairs of branes.
We propose  as well the construction of a non-BPS M10-brane from which the BPS M9-brane
is obtained after the tachyonic mode of an open M2-brane ending 
on it condenses. Therefore we are able to write the brane descent 
relations from M9, anti-M9 pairs in terms of M-theory 
spacetime-filling branes, thereby preserving the symmetries of the
eleven dimensional theory.

The study of 9-branes and 10-branes in M-theory should be carried out in the 
context of massive eleven dimensional supergravity, given that the
BPS M9-brane couples magnetically to the mass. It is well-known 
that a fully eleven dimensional Lorentz invariant massive 
supergravity cannot be constructed \cite{BDHS}. Nevertheless, the massive
Type IIA supergravity of Romans \cite{Romans} can be derived from
eleven dimensions if the condition of eleven dimensional Lorentz
invariance is relaxed and one assumes an isometric
eleventh direction that is gauged in the supergravity action
\cite{BLO}. In this paper, we will work in the framework of this massive 
eleven dimensional supergravity defined with a Killing isometry.
This theory is therefore only invariant under ten
dimensional Lorentz transformations. Accordingly, the branes living
in this massive eleven dimensional background are described by
worldvolume effective actions where the Killing direction is 
gauged\footnote{Also additional terms proportional to the mass are 
added \cite{BLO}.}. This is also
the case for the non-BPS M10-brane, as we will see.

The Born-Infeld part of the M9-brane effective action has been constructed
in \cite{BvdS, EL1}, and the Wess-Zumino term, which is the
relevant part for the study of brane solitons, has also been derived
recently in \cite{Sato}. This 
construction provides a description of the M9-brane in terms of
a gauged sigma-model\footnote{In such a way that the right number of 
degrees of freedom of a vector multiplet is recovered
\cite{BvdS}, and the brane consistently propagates in a massive
background.}.
The effective action of an M9, anti-M9 pair can be described by a 
similar but more complicated expression which contains as well 
explicit couplings to the tachyonic mode that must be present in this
unstable system. This tachyonic excitation must appear in the spectrum
of an open M2-brane stretched between the brane and the antibrane,
as predicted by the duality with the type IIA theory, where the open
string stretched between a D8, anti-D8 pair of branes contains this
type of excitation. The effective action of the M9, anti-M9 pair
can be derived by oxidating the effective action describing this
D8, anti-D8 pair\footnote{Effective actions describing type II
Dp, anti-Dp pairs of branes have been constructed in \cite{KW}.}.
Recalling that the brane antibrane system is characterised
by two field strengths (one for the brane and one for the antibrane)
and that the non-trivial character of 
the soliton can be carried by just one of the two,  
it is enough to look at the couplings in the worldvolume effective
action of the brane, in our case of the M9-brane, 
in order to find out the topological defects that the system 
supports, taking into account as well that for this analysis the
coupling of the tachyonic mode to the worldvolume effective action
does not play any role.

Keeping this in mind we analyse in section 2 the topological defects 
that can occur in the
worldvolume of $n$ M9, anti-M9 pairs of branes by studying the different
terms that couple in the WZ part of the worldvolume effective action
of a system of $n$ M9-branes. In this way we are able to derive
brane descent relations in M-theory from pairs of M9, anti-M9
branes. Then in section 3 we propose the construction of  a 
non-BPS, unstable, M10-brane from which the BPS M9-brane is obtained
through tachyonic condensation. Again, this tachyonic mode must be
associated to open M2-branes ending on the M10-brane, as required
by the duality with type IIA.
We thereby present the brane descent relations in terms of this 
spacetime-filling brane. In section 4 we discuss the existence of
other non-BPS unstable branes in M-theory and the possibility of
stabilising them within the Ho\u{r}ava-Witten construction \cite{HW}.
In particular, in this picture the tachyonic mode of the non-BPS 
M10-brane is projected out. Finally section 5 contains some conclusions.

\section{BPS M-branes as bound states}

In this section we show how the M-theory BPS branes can be realised as
bound states of $n$ (M9, anti-M9) pairs of branes. As we have mentioned
in the introduction we can ignore the couplings of the tachyon in
the worldvolume effective action of this system and set the field 
strength of the antibranes to zero, so that we can 
simply study the effective action of $n$ coincident M9-branes. 
We use the WZ term 
constructed recently in \cite{Sato}, but
only consider explicitly those couplings
that are relevant for the analysis of the topological defects. 
We also ignore all numerical prefactors and factors of $\alpha^\prime$
as well as the explicit couplings to the mass.

The WZ term of the M9-brane effective action contains the 
couplings\footnote{Eleven
dimensional fields are denoted with hats.} \cite{Sato}:

\begin{eqnarray}
\label{M9WZ}
&&S^{(M9)}_{\rm WZ}=\int_{R^{8+1}}\left[ i_{\hat k}{\hat B}^{(10)}+
i_{\hat k}{\hat N}^{(8)}\wedge {\hat {\cal F}}+i_{\hat k}{\hat {\tilde C}}
\wedge {\hat {\cal F}}\wedge {\hat {\cal F}}+ \right. \\ 
&&\left. +{\hat C}_{{\hat \mu}_1{\hat \mu}_2
{\hat \mu}_3}D{\hat X}^{{\hat \mu}_1}D{\hat X}^{{\hat \mu}_2}
D{\hat X}^{{\hat \mu}_3}\wedge
{\hat {\cal F}}\wedge {\hat {\cal F}}\wedge {\hat {\cal F}}+
{\hat A}\wedge {\hat {\cal F}}\wedge {\hat {\cal F}}
\wedge {\hat {\cal F}}\wedge {\hat {\cal F}}+\dots
\right]\, ,\nonumber
\end{eqnarray}

\noindent where the dots indicate the couplings that we have omitted.
This action is invariant under the local isometric
transformations generated by the Killing vector ${\hat k}^{\hat \mu}$,
since the pull-backs onto the worldvolume occur with covariant
derivatives (defined below) or contracting with the Killing vector.
This vector has to be taken along a worldvolume direction in order to
obtain the D8-brane after double dimensional reduction
(see \cite{BvdS}, \cite{EL1}). ${\hat B}^{(10)}$ is the spacetime field
electric-magnetic dual to the
mass, ${\hat N}^{(8)}$ is the dual of the Killing vector considered
as a 1-form \cite{BEL}, 
and ${\hat C}$, ${\hat {\tilde C}}$ are, respectively,
the 3- and 6-forms of eleven dimensional supergravity. ${\hat {\cal F}}$ 
is the field strength of the worldvolume vector field describing 
an M2-brane, wrapped on the Killing direction, 
ending on the brane: ${\hat {\cal F}}=d{\hat b}^{(1)}+i_{\hat k}{\hat C}
\equiv {\hat F}+i_{\hat k}{\hat C}$. Finally,
${\hat A}\equiv -{\hat k}^{-2}{\hat k}_{\hat \mu}\partial
{\hat X}^{\hat \mu}$, and the covariant derivatives are defined as:
$D{\hat X}^{\hat \mu}=\partial {\hat X}^{\hat \mu}+{\hat A}
{\hat k}^{\hat \mu}$ (see \cite{Sato} 
and \cite{BJO,BEL} for more details).

The M6-brane, or M-theory Kaluza-Klein monopole, is realised as a 
bound state (M9, anti-M9) \cite{HL1}. 
This can be simply read from the term:

\begin{equation}
\int_{R^{8+1}} i_{\hat k}{\hat N}^{(8)}\wedge {\hat F}
\end{equation}

\noindent in the M9-brane effective action. As mentioned in the
introduction, there must be a tachyonic mode in the spectrum of a 
wrapped M2-brane stretched between the M9 and the anti-M9,
whose condensation will be accompanied by a 
non-trivial magnetic flux, so that a coupling:

\begin{equation}
\int_{R^{6+1}} i_{\hat k}{\hat N}^{(8)}
\end{equation}

\noindent remains in the worldvolume. 
The M-theory Kaluza-Klein monopole is charged with
respect to this field \cite{BEL}, 
so this is the topological defect that remains
after the condensation.

The M5-brane is realised, in turn, as an instanton-like 
configuration\footnote{Since a single M9-brane contains in
its worldvolume a U(1) vector field a system of $n$ coincident
M9-branes is described by a U($n$) worldvolume gauge theory.
Therefore the corresponding effective action is given by the 
same expression (\ref{M9WZ}) with $F\in$ U($n$) and where the trace is
taken over the U($n$) indices.}.
The term:

\begin{equation}
\int_{R^{8+1}}(i_{\hat k}{\hat {\tilde C}})\wedge {\rm Tr}({\hat F}\wedge
{\hat F})
\end{equation}

\noindent in the worldvolume effective action of $n$ M9-branes
gives a coupling

\begin{equation}
\int_{R^{4+1}}(i_{\hat k}{\hat {\tilde C}})
\end{equation}

\noindent when
$\int_{R^4}{\rm Tr}({\hat F} \wedge {\hat F})={\rm integer}$,
i.e. when the homotopy group $\Pi_3(U(n))=Z$. This happens 
for all $n>2$, however in the particular
case $n=2^{k-1}$, where $2k$ is defined as the codimension of the
topological defect (in this case $k=2$), it is possible to give a 
representation of the tachyon vortex configuration (the generator of
$\Pi_{2k-1}(U(n))$) such that all higher and lower dimensional charges
vanish \cite{Witten}.
For our particular case $n=2$, and the 
M5-brane is realised as a bound state 
M5=2 (M9, anti-M9) \footnote{This definition
of the M5-brane seems to imply 
that a pile of coincident M5-branes would simply be described by a U($m$) gauge
theory. This does not contradict the fact that the M5-brane field
content must be that of the six dimensional antisymmetric tensor
multiplet, whose non-abelian extension is not known, because 
in the bound state construction the
M5-brane is wrapped on the Killing direction of the M9-brane
and therefore its field content must be that of the five dimensional vector
multiplet.}.

The term responsible for the realisation of the M2-brane as a 
topological defect is: 

\begin{equation}
\int_{R^{8+1}} {\hat C}_{{\hat \mu}_1{\hat \mu}_2{\hat \mu}_3}
D{\hat X}^{{\hat \mu}_1}D{\hat X}^{{\hat \mu}_2}D{\hat X}^{{\hat \mu}_3}
\wedge {\rm Tr}({\hat F}\wedge {\hat F}\wedge {\hat F})\, .
\end{equation}

\noindent Now we have $\Pi_5(U(4))=Z$, and the topological defect
is realised as M2=4 (M9, anti-M9), given that the non-trivial integration
of the gauge field gives rise to a coupling:

\begin{equation}
\int_{R^{2+1}}{\hat C}_{{\hat \mu}_1{\hat \mu}_2{\hat \mu}_3}
D{\hat X}^{{\hat \mu}_1}D{\hat X}^{{\hat \mu}_2}
D{\hat X}^{{\hat \mu}_3}\, ,
\end{equation}

\noindent which describes an M2-brane with an isometric transverse
direction\footnote{Since the M2-brane cannot move along this direction,
which plays the role of the eleventh coordinate,
it behaves effectively as a D2-brane, and a set of such coincident
M2-branes can then be described by a U($m$) gauge theory.}.

Finally, the term

\begin{equation}
\int_{R^{8+1}}{\hat A}\wedge {\rm Tr}({\hat F}\wedge {\hat F}\wedge
{\hat F}\wedge {\hat F})
\end{equation}

\noindent describes an M-wave as a bound state of 8 (M9, anti-M9) pairs.
In this case $\Pi_7(U(8))=Z$, and the field that remains after the
condensation:
${\hat A}=-{\hat k}^{-2}{\hat k}_{\hat \mu}\partial {\hat X}^{\hat \mu}$
is the field to which the M-wave couples minimally \cite{BT}.

These realisations of M-branes as bound states of M9, anti-M9 pairs 
are those that one would obtain by oxidising the 
brane descent relations in the Type IIA theory in terms of D8, 
anti-D8 pairs of branes. Here we have
shown that the M9-brane worldvolume effective action correctly contains
the couplings describing these realisations.
 
In the type IIA theory it is possible to write the brane descent
relations in terms of non-BPS spacetime-filling D9-branes, and in this
way all the symmetries of the spacetime are preserved \cite{Horava}.
This D9-brane is unstable because the open strings ending on it
contain tachyonic excitations. These modes however can condense in a 
kink (resp. anti-kink) configuration, giving rise to a D8 (resp. anti-D8) 
brane as the topological
defect. Oxidation to M-theory predicts then an unstable M10-brane
from which the M9-brane should be obtained after tachyonic 
condensation. In the next
section we analyse this possibility.

\section{The non-BPS M10-brane}

Having the action of the M9-brane it is
possible to construct the action of a non-BPS M10-brane which gives rise
to this brane after tachyonic condensation. In this case, since the
field strength ${\hat {\cal F}}$ is given by: ${\hat {\cal F}}=d{\hat b}^{(1)}+
i_{\hat k}{\hat C}$, the tachyonic mode must be associated to an 
M2-brane, wrapped
on the Killing direction, ending on the M10-brane. 

The clue for the construction of the effective action is to realise
that the M10-brane should reproduce
the Type IIA non-BPS D9-brane \cite{Horava} upon double dimensional 
reduction. The WZ term of the action of type II non-BPS D$p$-branes 
has been constructed in \cite{BCR}. In the particular case
of the D9-brane it reads\footnote{We
ignore the contribution of the A-roof genus and, again,
all numerical prefactors and factors of $\alpha^\prime$.}:

\begin{eqnarray}
\label{laD9}
S^{(D9)}_{\rm WZ}&=&\int_{R^{9+1}}\left[ C^{(9)}+C^{(7)}\wedge {\cal F}+
C^{(5)}\wedge {\cal F}\wedge {\cal F}+C^{(3)}\wedge {\cal F}
\wedge {\cal F}\wedge {\cal F}+\right.\nonumber\\
&&\left. +C^{(1)}\wedge {\cal F}\wedge {\cal F}\wedge {\cal F}
\wedge {\cal F}\right] \wedge dT\, ,
\end{eqnarray}

\noindent where $C^{(p)}$ denotes the $p$-form RR-potential,
${\cal F}=F+B^{(2)}$, with $F=db^{(1)}$ and $B^{(2)}$ the NS-NS 2-form,
and $T$ stands for 
the, real, tachyon field induced in the worldvolume by the open
strings ending on the brane. When the tachyon
condenses to a non-trivial kink configuration, depending on a single
coordinate $x$ and such that $\int dT(x)=\pm 2T_0$, where
$\{T_0,-T_0\}$ are the two minima of the tachyon potential, we
have, in the limit of zero size: $dT(x)=2T_0\delta (x-x_0)dx$,
and substituting in (\ref{laD9}) the effective action of a BPS D8-brane
localised in $x_0$ is obtained \cite{BCR}.

The WZ part of the effective action of the proposed non-BPS M10-brane is 
then given by\footnote{We use dots to denote those terms that
are not relevant for the analysis of the topological defects.}:

\begin{eqnarray}
\label{M10WZ}
&&S^{(M10)}_{\rm WZ}=\int_{R^{9+1}}\left[ i_{\hat k}{\hat B}^{(10)}+
i_{\hat k}{\hat N}^{(8)}\wedge {\hat {\cal F}}+i_{\hat k}{\hat {\tilde C}}
\wedge {\hat {\cal F}}\wedge {\hat {\cal F}}+ 
{\hat C}_{{\hat \mu}_1{\hat \mu}_2{\hat \mu}_3}.\right. \\ 
&&\left. .D{\hat X}^{{\hat \mu}_1}D{\hat X}^{{\hat \mu}_2}
D{\hat X}^{{\hat \mu}_3}\wedge
{\hat {\cal F}}\wedge {\hat {\cal F}}\wedge {\hat {\cal F}}+
{\hat A}\wedge {\hat {\cal F}}\wedge {\hat {\cal F}}
\wedge {\hat {\cal F}}\wedge {\hat {\cal F}}+\dots \right]\wedge d{\hat T}\, ,
\nonumber
\end{eqnarray}

\noindent since this action reproduces (\ref{laD9}) after double
dimensional reduction along the Killing direction. 
Also, the condensation of the tachyon field
into a non-trivial kink configuration, 
$d{\hat T}(x)=2{\hat T}_0 \delta (x-x_0) dx$, in the limit of zero
size, gives the M9-brane effective action that we
considered in the previous section. 
Like in the non-BPS D9-brane, the tachyon field living in the worldvolume
of the M10-brane is a real scalar, in this case induced by open M2-branes wrapped
on the Killing direction\footnote{The presence
of a tachyonic mode in the spectrum of open M2-branes
ending on the M10-brane is inferred by duality with the type IIA
theory, but being this a strong-weak coupling duality the open M2-branes
are intrinsically non-perturbative and the presence of this
instability cannot be tested by any perturbative methods.}.

Regarding the DBI part of the effective action, it has been argued in
\cite{Sen2} that non-BPS Dp-branes may be described by a DBI 
action\footnote{Here we are only concerned with the bosonic part of
the action. See \cite{Sen2} for more details.}:

\begin{equation}
\label{DpDBI}
S^{({\rm Dp})}_{{\rm DBI}}=-\int_{R^{p+1}} \sqrt{|{\rm det}(G+{\cal F})|}
\,R(T,\partial T,\dots)\, ,
\end{equation}

\noindent where $R$ is some function of the tachyon field vanishing
at the minimum of the tachyon potential. In this way
the worldvolume action vanishes
identically, and can describe a configuration indistinguishable from
the vacuum. On the other hand, for vanishing tachyon field $R$ gives
a constant, and the action in \cite{Sen2} for a non-BPS Dp-brane 
involving the massless fields is recovered. 
For a tachyonic kink configuration in the zero size limit
it has been argued that
$R(x)\sim C \delta(x-x_0)$ \cite{Kluson}, and (\ref{DpDBI})
reduces to the DBI part of the effective action of a BPS 
D(p-1)-brane.

Similarly, we can argue that the DBI part of the non-BPS
M10-brane effective action is given by: 

\begin{equation}
\label{M10DBI}
S^{({\rm M10})}_{{\rm DBI}} =-\int_{R^{9+1}} |{\hat k}|^3 
\sqrt{|{\rm det} \left( {\hat \Pi}+ |{\hat k}|^{-1}  
{\hat {\cal F}} \right)|}
\,{\hat R}({\hat T},\partial {\hat T},\dots)\, ,
\end{equation}

\noindent such that when the tachyon condenses to its minimum
the DBI effective action
of the M9-brane is recovered \cite{BvdS, EL1}:

\begin{equation}
\label{M9DBI}
S^{({\rm M9})}_{{\rm DBI}}=-\int_{R^{8+1}}|{\hat k}|^3 
\sqrt{|{\rm det} \left( {\hat \Pi}+ |{\hat k}|^{-1}
{\hat {\cal F}} \right)|}\, .
\end{equation}

\noindent Here ${\hat \Pi}$ is the pull-back of the spacetime
metric:

\begin{equation}
{\hat \Pi} = D {\hat X}^{\hat \mu} D {\hat X}^{\hat \nu} 
{\hat g}_{{\hat \mu}{\hat \nu}} = 
\partial {\hat X}^{\hat \mu} \partial {\hat X}^{\hat \nu} 
\left( {\hat g}_{{\hat \mu}{\hat \nu}}
+ |{\hat k}|^{-2}{\hat k}_{\hat \mu}{\hat k}_{\hat \nu} \right) \, .
\end{equation}

The proposed M10-brane 
action, given by (\ref{M10WZ}) and (\ref{M10DBI}), contains a Killing direction
in its worldvolume. Therefore it is only invariant under ten dimensional
Lorentz transformations, consistently with the fact that it
is defined in a massive eleven dimensional background.
In this sense the M10-brane that we have constructed
is spacetime-filling and preserves all the symmetries of the massive
eleven dimensional spacetime, including the invariance under local
isometric transformations.

Finally, since both the M9 and the anti-M9 branes can be obtained
from the M10-brane when the tachyon condenses to a kink or an
anti-kink configuration \cite{Horava}, we
can write the brane descent relations that we derived in the previous
section in terms of spacetime-filling
branes as: M9=M10, M6=2 M10, M5=4 M10, M2=8 M10, 
M-wave=16 M10 \footnote{The WZ part
of the effective action describing the set of 
$n$ coincident M10-branes
is of the same form (\ref{M10WZ}) but now ${\hat F}\in$ U($n$), the
tachyon transforms in the adjoint representation \cite{Horava} 
and traces over the U($n$) indices are taken.}.

\section{Other non-BPS branes in M-theory}

We have seen how the duality between M-theory and type IIA predicts
the existence of an unstable non-BPS M10-brane from which the type IIA
D9-brane is obtained after double dimensional reduction. Similarly, it
is possible to construct unstable non-BPS branes in M-theory giving
rise to the whole spectrum of non-BPS D-branes in type IIA. One finds
that apart from the M10-brane there are M(-1), M1, M4, M5 and M8
non-BPS branes\footnote{Among these branes, the M(-1), M1, M4 and M5 can 
exist as well in massless M-theory.}. 
These branes can be described as bound states of
BPS M-branes as: Mp=(M(p+1), anti-M(p+1)). In this brane antibrane
unstable system the complex tachyonic excitation in the
open M2-branes stretched between the brane and the antibrane
condenses through a kink configuration, giving rise to the non-BPS
M-brane, which is also unstable, because the open M2-branes ending on 
it contain still a real tachyonic excitation.

The tachyonic excitation in the non-BPS Mp-brane can in turn
condense, and give rise to a BPS M-brane. A careful analysis shows
that the dimension of this resulting M-brane  
depends on whether the M2-brane ending on the non-BPS Mp-brane,
whose tachyonic mode is condensing, is wrapped or unwrapped 
\footnote{ As explained in \cite{HL1} certain realisations of branes
as solitons in brane anti-brane systems require to choose one
special direction.}. 
For the M1, M5 and M8 branes
the tachyonic mode is associated to a wrapped M2-brane ending on the
brane, and one can see that the BPS object that remains after the
condensation must be an M0, a wrapped M5 and an M6 brane respectively. 
In the case of the non-BPS M4-brane the tachyonic mode is associated
to unwrapped M2-branes ending on the brane, and the object that remains after
the condensation is an M2-brane.
 
Thus, the existence of these non-BPS branes in 
M-theory allows the construction of
BPS branes from brane antibrane pairs in two steps. First, the complex
tachyonic mode of the brane stretched between the brane and the antibrane
condenses, giving rise to a non-BPS M-brane, with a real tachyonic mode
associated to open M2-branes ending on it. Second, this tachyonic mode
condenses in the same type of configuration and a stable BPS M-brane 
emerges. This generalises to M-theory the two-step construction of type II 
theories of \cite{Sen1,Sen3}. 

It is interesting to note that the non-BPS M-branes that we have
considered give rise, upon reduction, not only to non-BPS 
Dp-branes but also to some additional non-BPS branes in
the type IIA theory. These branes predict in turn the existence of
similar non-BPS branes in type IIB by T-duality. We have mentioned in the
introduction that
BPS NS-NS and gravitational branes \cite{Hull} in the type IIA and
type IIB theories can be interpreted as solitons in brane antibrane
systems \cite{Yi, HL1}, with tachyonic condensing charged
objects that are extended and non-perturbative. In the realisations 
discussed in these references the complex tachyonic
mode condenses in a vortex-like configuration, and the resulting 
solitonic object is stable and carries NS-NS or gravitational charge.
One could instead consider a two-step construction of these BPS branes
as in \cite{Sen1,Sen3}, and this would lead to new 
intermediate non-BPS ``NS-NS'' and ``gravitational''
branes\footnote{NS-NS and gravitational in the sense that they are
derived in the two-step construction of this kind of branes, but
it is clear that they do not carry any charge.} 
as remnants of tachyonic kink configurations. These are the 
additional non-BPS branes that are derived from M-theory. 

\subsection{Stable non-BPS M-branes in the Ho\u{r}ava-Witten construction}

Unstable non-BPS branes can become stable when
the theory is orbifolded/ orientifolded by an appropriate 
symmetry (see \cite{reva}). 
Therefore, it is interesting to analyse whether the 
unstable non-BPS M-branes that we have discussed can become stable in
the framework of the Ho\u{r}ava-Witten construction \cite{HW}.
One can predict stable non-BPS
branes in this construction by uplifting the stable non-BPS branes of
the type I' theory, which have in turn been analysed in \cite{BGH} 
by using the T-duality connection with type I.

The spectrum of stable non-BPS type I' branes consists on a D(-1),
D0, D1, D6, D7, D8 and D9 branes, where the D0, D1, D8 and D9 are
stretched in the interval\footnote{The D6, D7, D8 and D9
branes are however unstable due to the presence of a tachyonic mode
in the open strings with one end on the brane and the other in one
of the 16 + 16 D8-branes located on top of the orientifold
fixed planes. This can be inferred from the type I
case \cite{FGLS}.}.
These branes can be constructed as bound
states: D(-1)=(D0, anti-D0), D0=(D0, anti-D0) (in this case the
D0's are `stretched'), D1=(D2, anti-D2) (stretched),
D6=(D6, anti-D6), D7=(D8, anti-D8), D8=(D8, anti-D8) (stretched).  
Uplifting these configurations
to M-theory we find: \\
M(-1)=(M0, anti-M0), M0=(M0, anti-M0) 
(`stretched'), M1=(M2, anti-M2) (stretched), M6=(M6, anti-M6),
M8=(M9, anti-M9), M9=(M9, anti-M9) (stretched), together with a
stretched M10-brane. 

We can thus conclude that the non-BPS M(-1), M1, M8 and M10 
branes of uncompactified M-theory are
stabilised when one considers M-theory in the interval. Moreover, we
find additional branes realised as BPS brane
antibrane pairs in which the tachyonic mode is removed from the
spectrum by the orbifold projection, and no condensation occurs.
This happens for the M0, M6 and M9 non-BPS branes.

The Ho\u{r}ava-Witten construction can be described 
as M-theory orientifolded by $I_{10}{\hat \Omega}$ (see \cite{BEHHLvdS}),
where $I_{10}: x^{10} \rightarrow -x^{10}$ and 
${\hat \Omega}$ is the operation that reverses the orientation
of the M2 and the M5 branes. 
In this description
the end of the world branes of Ho\u{r}ava-Witten are identified as
the two fixed planes associated to the orientifold projection with
16 M9-branes on top of them. The connection with the
type I' theory through compactification on $S^1$ implies that the
M6, M8, M9 and M10 branes must become unstable due to the presence of
a tachyonic mode in the open M2-branes with one end on the brane
and the other in one of the 16 + 16 M9-branes located on the top of the 
orientifold fixed planes.

\section{Conclusions}

We have shown how the brane descent relations that one expects to find
in M-theory by oxidation from Type IIA are indeed predicted
by the analysis of the topological defects that can occur in the worldvolume
of $n$ M9, anti-M9 pairs of branes. We have proposed as well a
non-BPS, unstable, M10-brane, from which the brane descent
relations can be expressed in terms of spacetime-filling branes, 
preserving all the 
symmetries of the spacetime. Since massive eleven dimensional
supergravity is at present only known explicitly when the spacetime
contains a Killing isometry \cite{BLO}, the corresponding 
M10-brane contains as
well this Killing isometry in its worldvolume, and therefore 
it is not fully eleven dimensional Lorentz
invariant, consistently with the type of eleven
dimensional spacetime in which it is defined.

Using the relation between ordinary and higher K-theory
groups (see \cite{Horava}, \cite{OS}): 
${\tilde K}(S^{10-p})=K^{-1}(S^{9-p})$ , 
it is inferred that the branes that can be obtained in the
descent construction in this massive M-theory  are
classified according to:

\begin{displaymath}
{\tilde K}(S^{10-p}) =
\mbox{ 
$\left\{ \begin{array}{ccccc}
         Z & {\rm for} & p  &  {\rm even}  \\
         0 & {\rm for} & p & {\rm odd}           
                                    \end{array} \right.$\, .}   
\end{displaymath}

\noindent Therefore the K-theory groups predict the M-wave, 
M2 and M6 branes\footnote{For the M-wave and the M6 
the $S^1$ direction coincides with their
special isometric direction.}, and an M4 and M8 branes, which are
identified as wrapped M5 and M9 branes, identification that is
supported by the bound state analysis that we have carried out in
this paper.

We have discussed as well the existence of other non-BPS branes in
M-theory, giving rise to type IIA non-BPS D-branes
after dimensional reduction. These M-branes predict unstable non-BPS
``NS-NS'' and ``gravitational'' branes in both the type IIA and type IIB
theories, which would arise naturally in a two-step construction, as
in \cite{Sen1,Sen3}, of NS-NS and gravitational BPS branes. 

Finally, we have shown that when M-theory is compactified in the
interval some of the non-BPS M-branes become stable, in particular
the spacetime-filling M10-brane, since they give
rise to stable non-BPS branes in type I' after reduction. The
K-theory classification of stable M-branes in the Ho\u{r}ava-Witten
picture seems to be given by the K-theory group:
$KR(S^{9-p}\times S^{1,1}, S^{1,1})$, which provides 
the M-theory interpretation of the type I' K-theory group
\cite{Witten,Horava,BGH}: 
$KR^{-1}(S^{8-p} \times S^{1,1},S^{1,1})$.
Some of the non-BPS branes predicted by this group become however
unstable due to the presence of background M9-branes in the 
Ho\u{r}ava-Witten picture, which, as in the type I case \cite{FGLS}, 
do not play any role in the K-theory classification.

\subsection*{Acknowledgements}

L.~H. would like to acknowledge the support of the European Commission 
TMR programme grant ERBFMBICT-98-2872, and the Theory Division at CERN,
where this work has been completed, for hospitality. The I.C. Theory
Group is supported by PPARC under SPG grant 613.

\end{document}